\documentclass[12pt]{article}
\usepackage{cite,graphicx,amsmath,amssymb}
\usepackage{graphicx}
\usepackage{multicol}
\usepackage{xcolor}

\newcommand{\be}{\begin{equation}}
\newcommand{\ee}{\end{equation}}
\newcommand{\bea}{\begin{eqnarray}}
\newcommand{\eea}{\end{eqnarray}}
\begin{document}

\thispagestyle{empty}
\vspace*{.2cm}
\noindent
HD-THEP-10-14 \hfill 22 July 2010

\vspace*{2.0cm}

\begin{center}
{\Large\bf Precision Gauge Unification\\[.3cm] from Extra Yukawa Couplings}
\\[2.5cm]

{\large Ivan Donkin and Arthur Hebecker\\[6mm]}

{\it
Institut f\"ur Theoretische Physik, Universit\"at Heidelberg, 
Philosophenweg 19, D-69120 Heidelberg, Germany\\[3mm]
{\small\tt (\,a.hebecker@thphys.uni-heidelberg.de} {\small ,} 
{\small\tt \,i.donkin@thphys.uni-heidelberg.de}\small\tt \,) }
\\[2.0cm]

{\bf Abstract}
\end{center} 

\noindent
We investigate the impact of extra vector-like GUT multiplets on the 
predicted value of the strong coupling. We find in particular that 
Yukawa couplings between such extra multiplets and the MSSM Higgs doublets 
can resolve the familiar two-loop discrepancy between the SUSY GUT prediction 
and the measured value of $\alpha_3$. Our analysis highlights the advantages 
of the holomorphic scheme, where the perturbative running of gauge couplings 
is saturated at one loop and further corrections are conveniently described 
in terms of wavefunction renormalization factors. If the gauge couplings as 
well as the extra Yukawas are of ${\cal O}(1)$ at the unification scale, 
the relevant two-loop correction can be obtained analytically. However, the 
effect persists also in the weakly-coupled domain, where possible 
non-perturbative corrections at the GUT scale are under better control.

\newpage
\section{Introduction}

The consistency of low-energy data with supersymmetric gauge coupling 
unification \cite{Dimopoulos:1981yj} is one of the strongest reasons 
to expect the discovery of supersymmetry at the LHC. Moreover, gauge 
coupling unification is very well-motivated in heterotic string 
compactifications (see \cite{Buchmuller:2005jr} for some of the recent 
developments) as well as in F-theory \cite{Beasley:2008dc}. 
Thus, if supersymmetry is discovered, the SUSY GUT framework will provide 
one of the most direct ways to access the fundamental high-scale theory
(see \cite{Adam:2010uz} for a recent phenomenological study).

The SUSY GUT prediction for the strong coupling is, however, not perfect. 
In fact, using two-loop $\beta$ functions and identifying the effective 
SUSY breaking scale with $m_Z$, the prediction misses the measured value 
of $\alpha_3(m_Z)$ by many standard deviations. Possible resolutions of 
this discrepancy include corrections due to an unusual SUSY spectrum (for 
a recent analysis see \cite{Raby:2009sf}) or unexpectedly large GUT 
thresholds.

In the present paper, we focus on a different possibility: In addition to 
the usual MSSM spectrum, we allow for extra chiral supermultiplets in 
complete vector-like GUT representations. Such states appear in many 
different contexts and are well-motivated theoretically (e.g. as the 
``messengers'' of gauge mediation). While these extra multiplets do 
not affect the one-loop prediction for $\alpha_3$, the induced two-loop 
effect is significant. Unfortunately, it further enhances the familiar 
problems of the MSSM two-loop prediction \cite{AmelinoCamelia:1998tm}.

However, with extra multiplets naturally come extra Yukawas. We focus on
Yukawa couplings of the extra multiplets with the MSSM Higgs doublets (in
line with the general structure of the MSSM and with the natural extension of
R-parity to these multiplets). As a simple example one may think of a fourth
vector-like generation. We also note that, following \cite{Moroi:1991mg,
Tobe:2002zj,Babu:2004xg}, a similar class of models has recently been 
considered in \cite{Babu:2008ge,Graham:2009gy,Martin:2010dc} as a possible 
solution to the little hierarchy problem. Furthermore, extra multiplets
may also improve the little hierarchy problem in the context of 
\cite{Barbieri:2007tu}. We take all of this as an important 
extra motivation for our scenario. 

Our results show that the new Yukawa interactions induce a significant 
shift in the strong coupling towards its correct experimental value. Moreover,
we find that if the extra Yukawas are relatively large at the GUT scale, we 
end up with an almost perfect prediction for $\alpha_3$. 

The paper is organized as follows: We start in Sect.~\ref{cu} with a two-loop 
analysis of MSSM gauge unification in the holomorphic scheme. While gauge 
couplings run only at one loop, the $Z$ factors of chiral multiplets receive 
corrections at all loop orders. However, to achieve two-loop precision for 
the $\alpha_3$ prediction, it is sufficient work with one-loop $Z$ factors. 
Their effect on the $\alpha_3$ prediction comes from the transition to the 
canonical scheme (via the vector and Konishi anomalies), which we perform at 
the electroweak scale. In this approach, the two-loop correction to the MSSM 
prediction for $\alpha_3$ arises from a sum of terms $\sim \ln Z_f$, where 
$f$ runs over all flavors, including in particular the two Higgs doublets. 
It becomes apparent that a significant enhancement of the Higgs $Z$ factors 
can provide the desired shift in the $\alpha_3$ prediction.

Extra multiplets are introduced in Sect.~\ref{noyukawa}. Their $Z$ factors are
not important since these fields are integrated out above the scale where 
the transition to the canonical scheme is performed. The detrimental effect 
of extra multiplets on gauge unification mentioned earlier arises solely 
through the increased value of the gauge couplings at high energies. The 
gauge couplings lead to decreasing $Z$ factors as one runs from high to low 
energy scales, and this effect is enhanced in the presence of extra matter.

In Sect.~\ref{strong} we introduce extra Yukawa couplings. We first consider an 
analytically calculable example with strong gauge coupling at the GUT scale. 
This can be realized introducing extra multiplets in the ${\bf 10} +
\overline{\bf 10}$ and ${\bf 5} + \overline{\bf 5}$ of SU(5) with masses near 
the TeV scale. Assuming that the extra Yukawas of type ${\bf 10}\,{\bf 10}\,H_U$
and $\overline{\bf 10}\,\overline{\bf 10}\,H_D$ are also strong at the GUT 
scale and neglecting the small effect of the top Yukawa coupling, we solve 
this model analytically. The resulting shift in the $\alpha_3$ prediction, 
which is due to enhanced Higgs $Z$ factors, leads to nearly perfect 
agreement with the experimental value. 

In Sect.~\ref{perturbative} we extend our analysis to more general scenarios, 
keeping in particular the GUT-scale gauge coupling in the perturbative domain. 
We demonstrate that our promising initial results retain their validity in 
this setting as long as the extra Yukawa couplings at the unification scale 
are sufficiently large. 

Scenarios with a larger number of extra Yukawa couplings are investigated in
Sect.~\ref{extended}. We focus on two particular types of extension. First we
analyze models in which  further $\mathbf{10}+\overline{\mathbf{10}}$  pairs
with further extra Yukawas are introduced. We find that in this case the total
effect on the Higgs $Z$ factor can not be increased significantly. The
second type of models we consider possess the same matter content as our
minimal model from Sect.~\ref{perturbative} (i.e. a $\mathbf{10}+
\overline{\mathbf{10}}$ plus a $\mathbf{5}+\overline{\mathbf{5}}$ pair).
However, we now also allow for couplings of  type $\overline{\bf 5}\,{\bf 10}\,H
_D$
and ${\bf 5}\,\overline{\bf 10}\,H_U$. We find that in this case the two-loop
prediction for the inverse coupling $2\pi/\alpha_3$ is increased even further.
In particular, we are able to reproduce the `optimal' results from
Sect.~\ref{perturbative} under milder assumptions (i.e. for lower couplings at
the GUT scale).

Low- and high-scale threshold corrections and, in particular, the critical 
issue of strong gauge couplings at the GUT scale \cite{Moroi:1993zj,
Kolda:1996ea,Ghilencea:1997mu,Ghilencea:1997yr,Jones:2008ib,Kopp:2009xt,
Sato:2009yt,Abel:2008tx} are discussed in Sect.~\ref{thresh}. We emphasize 
that, due to the absence of higher-order perturbative corrections to the 
holomorphic couplings, precision unification is not compromised when 
allowing for relatively large values of the GUT coupling. However, once 
one reaches the actual strong-coupling regime, non-perturbative corrections 
can arise and affect the $\alpha_3$ prediction. Thus, it appears to be safe 
to stay in a domain where terms that are exponentially suppressed by the 
inverse gauge coupling are negligible. This is sufficient for our purposes.

\section{Conventional unification from a \\ holomorphic perspective}
\label{cu}

\subsection{Basic formulae with holomorphic thresholds}
\label{holomorphic}

We find it convenient to work in a holomorphic scheme, where the 
perturbative running of gauge couplings is saturated at one loop \cite{
Novikov:1983uc,Shifman:1996iy,ArkaniHamed:1997mj}. Focussing for simplicity 
on SU(5) models with adjoint breaking, we have
\be
\frac{2\pi}{\alpha_{h,i}(\mu)}=\frac{2\pi}{\alpha_h(M)}
+b_i\ln\frac{M}{\mu}+b_i^{(3)}\ln\frac{M}{m_3^h}+b_i^{X,Y}\ln\frac{M}
{m_{X,Y}^h}+b_i^\Phi\ln\frac{M}{m_\Phi^h}\,,\label{holmaster}
\ee
where $\alpha_{h,i}(\mu)$, with $i=1,2,3$, are the holomorphic U(1), 
SU(2) and SU(3) gauge couplings of the MSSM at some low scale $\mu$ and 
$\alpha_h(M)$ is the unified holomorphic coupling at some high scale $M$. 
This equation is exact to all orders of perturbation theory provided 
that $\mu$ lies above the supersymmetry breaking scale.
While the first logarithm combines the effects of all fields that remain 
light below the GUT scale, the last three logarithms are associated with 
heavy fields. The various $b_i$'s are the appropriate one-loop $\beta$-function
coefficients. They are labelled by `(3)' for Higgs-triplet, `$X,Y$'
for the massive vector multiplets of $X,Y$ gauge bosons, and $\Phi$ for 
those components of the GUT-breaking field $\Phi$ that are not `eaten' by 
the $X,Y$ bosons. We emphasize that $m_3^h$, $m_{X,Y}^h$ and $m_\Phi^h$ 
are holomorphic rather than physical mass scales. In other words, these 
are the mass parameters of the holomorphic Wilsonian action, in which 
kinetic terms are {\it not} canonically normalized \cite{ArkaniHamed:1997mj}. 

At the moment, we may think of a situation where $m_3^h\sim m_{X,Y}^h \sim 
m_\Phi^h \sim M^h_{\rm GUT}$ and $\mu\ll M^h_{\rm GUT}\ll M$. Furthermore, 
we identify the SUSY-breaking scale with $m_Z$ for now, postponing a brief
discussion of low-scale threshold effects to Sect.~\ref{thresh}.

Thus, we can set $\mu=m_Z$ and translate holomorphic to canonical 
gauge couplings using the well-known anomaly relation \cite{Novikov:1983uc,
Shifman:1996iy,ArkaniHamed:1997mj}\footnote{At our level of accuracy, we
can ignore corrections associated with the transition to the scheme-dependent
(e.g. DRED) physical gauge coupling \cite{Jack:1996cn}.}:
\be
\frac{2\pi}{\alpha_{h,i}}=\frac{2\pi}{\alpha_i}+T(G_i)\ln
g_i^2+\sum_f T(R_i^f)\ln Z_f\,.\label{anom}
\ee
Here $\alpha_i\equiv g_i^2/(4\pi)$ is the canonical gauge coupling,
$T(G_i)=C_2(G_i)$ is the Dynkin index or quadratic Casimir of the
adjoint of the gauge group $G_i$, and $T(R_i^f)$ is the Dynkin index of 
the representation $R_i^f$ of the flavor $f$ (with respect to $G_i$).
This gives rise to the `holomorphic master formula'\footnote{
i.e.
the holomorphic version of what is called the `master formula' in 
\cite{Shifman:1996iy,Ghilencea:1997mu}
}
\bea
\frac{2\pi}{\alpha_i(m_Z)}&=&\frac{2\pi}{\alpha_h(M)}+b_i\ln\frac{M}{m_Z}+
b_i^{(3)}\ln\frac{M}{m_3^h}+b_i^{X,Y}\ln\frac{M}{m^h_{X,Y}}
+b_i^\Phi\ln\frac{M}{m^h_\Phi}\nonumber \\
\label{hmf}\\
&&-T(G_i)\ln(g_i^2(m_Z))-\sum_f T(R_i^f)\ln Z_f(m_Z)\,.
\nonumber
\eea
Obviously, the choice of $M$ in this formula is irrelevant for the 
prediction of $\alpha_3$ from $\alpha_1$ and $\alpha_2$. In particular,
we can set $M=m^h_{X,Y}$ (irrespective of the actual UV completion scale) and 
write
\be 
\frac{2\pi}{\alpha_i(m_Z)} = \frac{2\pi}{\alpha_h(m^h_{X,Y})}+
b_i\ln\frac{m^h_{X,Y}}{m_Z}-T(G_i)\ln(g_i^2(m_Z))-\sum_f 
T(R_i^f)\ln Z_f(m_Z)+\Delta^h_{i,{\rm GUT}}\,,\label{aih}
\ee
with (holomorphic) GUT threshold corrections 
\be
\Delta^h_{i, {\rm GUT}} = 
b_i^{(3)}\ln\frac{m^h_{X,Y}}{m^h_3}+b_i^\Phi\ln\frac{m^h_{X,Y}}{m^h_\Phi}\,.
\label{dih}
\ee

The above threshold corrections will be small if the two relevant mass
ratios are ${\cal O}(1)$. This will be the case if the superpotential 
contains no parametrically small couplings. Indeed, staying strictly 
within the holomorphic scheme, not only $m^h_\Phi$ and $m^h_3$, but also 
$m^h_{X,Y}$ are determined purely by superpotential terms (in contrast to 
the physical masses of the $X,Y$ multiplets, which include a prefactor 
$g$ coming form the gauge kinetic term).

\subsection{Compatibility with the conventional master formula}
\label{canonical}
In order to establish the equivalence between Eq.~(\ref{hmf}) and the 
more conventional master formula of \cite{Shifman:1996iy}, we 
need to replace holomorphic by physical mass parameters: 
\be
m^p_{X,Y} = g\, m^h_{X,Y} Z^{1/2}_\Phi \,, \quad\quad 
m^p_{\Phi}Z_\Phi = m^h_{\Phi} \,, \quad\quad
m^p_{3}Z_3 = m^h_3\,,  \label{phvshol}
\ee
where $g$, $Z_\Phi$ and $Z_3$ are the (GUT-scale) gauge coupling and $Z$
factors\footnote{
Since they appear only in logarithms, it does not matter whether we use 
a holomorphic or a canonical gauge coupling and at which precise mass scale
we evaluate all these quantities.
} 
and the superscript `$p$' stands for `physical'. Using these relations 
together with Eq.~(\ref{anom}), we rewrite Eq.~(\ref{hmf}) in terms of 
canonical gauge coupling and physical masses. Making also use of the 
identity $b_i^{X,Y}-2T(G_i)=-2T({\rm SU(5)})$, which follows from 
$b_i^{X,Y}=-2T(R_i^{X,Y})$ and $T(R_i^{X,Y})+T(G_i)=T({\rm SU(5)})$,
and choosing $M=m^p_{X,Y}$, we find
\be
\frac{2\pi}{\alpha_i(m_Z)} = \frac{2\pi}{\alpha(m^p_{X,Y})}+
b_i\ln\frac{m^p_{X,Y}}{m_Z}-T(G_i)\ln\frac{\alpha_i(m_Z)}
{\alpha(m^p_{X,Y})}-\sum_f T(R_i^f)\ln\frac{Z_f(m_Z)}{Z_f(m^p_{X,Y})}+
\Delta^p_{i,{\rm GUT}}\,,
\label{aip}
\ee
with (physical) GUT threshold corrections 
\be
\Delta^p_{i, {\rm GUT}} = 
b_i^{(3)}\ln\frac{m^p_{X,Y}}{m^p_3}+b_i^\Phi\ln\frac{m^p_{X,Y}}{m^p_\Phi}\,.
\label{dip}
\ee

This is in agreement with \cite{AmelinoCamelia:1998tm, 
Ghilencea:1997mu,Shifman:1996iy}. It simply represents a slightly different 
parameterization of our ignorance of the high-scale input: In the 
`holomorphic master formula' of the previous subsection, all the relevant 
non-holomorphic (and in this sense `unprotected') input data were the 
high-scale boundary values of the Higgs $Z$ factors. (Note that high-scale
 $Z$ factors of complete SU(5) multiplets, such as MSSM matter, do not 
affect the $\alpha_3$ prediction.) Now, by contrast, the very same ignorance 
is hidden in the value of the physical Higgs triplet mass.

\subsection{Predicting $\alpha_3$}\label{prediction}

It will be convenient to rewrite Eqs.~(\ref{aih}) and (\ref{dih}) as:
\be 
\frac{2\pi}{\alpha_i(m_Z)} = \frac{2\pi}{\alpha_h(m^h_{X,Y})}+
b_i\ln\frac{m^h_{X,Y}}{m_Z}+\Delta_i\,\label{compact},
\ee
collecting all two-loop effects (the logs of gauge couplings and 
$Z$-factors as well as GUT thresholds) in the correction terms $\Delta_i$. 
Multiplying the first of these three equations by $(b_2-b_3)/(b_1-b_2)=5/7$, 
the second by $(b_3-b_1)/(b_1-b_2)=-12/7$, and adding them to the third 
equation, one finds
\be
\frac{2\pi}{\alpha_3} = 
\underbrace{\left[-\frac{5}{7}\,\frac{2\pi}{\alpha_1}+\frac{12}{7}\,
\frac{2\pi}{\alpha_2}\right]}_{\rm one-loop \,\, prediction}\,\,+\,\,
\underbrace{\left[\,\frac{5}{7}\, \Delta_1-\frac{12}{7} \, \Delta_2 
\, + \, \Delta_3\right]}_{\rm two-loop \,\, corrections}\,.\label{a3}
\ee

Here we have suppressed the mass scale, $m_Z$, for brevity. Setting 
the holomorphic GUT thresholds to zero, the two-loop correction, 
i.e. the second bracket in Eq.~(\ref{a3}), explicitly reads
\bea
\Delta_{\rm 2-loop}\left(\frac{2\pi}{\alpha_3}\right) &=&
\frac{24}{7}\ln g_2^2-3\ln g_3^2 + \frac{27}{14}\ln Z_L -
\frac{9}{7}\ln Z_E + \frac{9}{2}\ln Z_Q \label{da3}\\
&&- \frac{45}{14}\ln Z_U -\frac{27}{14}\ln Z_D +
\frac{9}{14}\ln Z_{H_U} +\frac{9}{14}\ln Z_{H_D}\,.
\nonumber
\eea
Given that we are only aiming at two-loop accuracy, it is consistent 
to evaluate all quantities entering the above expression  with 
one-loop precision. In particular, we can use one-loop $Z$-factors 
and the experimental values for the gauge couplings (given by
$2\pi/\alpha_1 = 370.7$, $2\pi/\alpha_2 = 185.8$, $2\pi/\alpha_3 = 
53.2$). The $Z$ factors can be written as \cite{Shifman:1996iy}
\bea
Z_{L,H_U,H_D} &=& Z_{SU(2)} \times Z_{U(1)}\nonumber \\
Z_Q &=& Z_{SU(3)} \times Z_{SU(2)} \times Z_{U(1)}^{1/9}\nonumber \\
Z_E &=& Z_{U(1)}^{4}   \label{zg} \\
Z_U &=& Z_{SU(3)}  \times Z_{U(1)}^{16/9}\nonumber \\
Z_D &=& Z_{SU(3)}  \times Z_{U(1)}^{4/9}\,. \nonumber 
\eea
Introducing the shorthand notation $Z_{U(1)}\equiv Z_1$, $\,\,Z_{SU(2)}\equiv 
Z_2$, $\,\,Z_{SU(3)}\equiv Z_3$, the respective gauge group contributions read
\bea
Z_{i} &=& 
\left(\frac{\alpha_{\rm GUT}}{\alpha_i(m_Z)}\right)^{-\frac{2C_{F,i}}{b_i}} \,.
\label{zf}
\eea
Here $C_{F,i}$ are the quadratic Casimir operators of the fundamental 
representations of U(1), SU(2) and SU(3), whereas $b_i$ are 
the respective one-loop $\beta$-function coefficients (cf. App.~A). 
Furthermore, $\alpha_{\rm GUT}$ is the one-loop  GUT coupling, 
$2\pi/\alpha_{\rm GUT}=153$. Equation (\ref{zf}) is valid under
the assumption that all $Z$-factors are  unity at $M_{\rm GUT}$.

Combining all contributions, we find:
\bea
\frac{2\pi}{\alpha_3(m_Z)}  &=&  \underbrace{53.7}_{\rm 1-loop}  - 
\,\underbrace{4.08}_{\rm vector\,\, anomaly} - 
\underbrace{0.64}_{Z_L} - \underbrace{5.51}_{Z_Q}  + 
\underbrace{0.21}_{Z_E} + \underbrace{3.22}_{Z_U}  + 
\underbrace{1.83}_{Z_D} - \underbrace{0.42}_{Z_{H_U},\,Z_{H_D}}
\nonumber \\ \nonumber \\
&=&  \, 53.7 \, - 5.4 \, = \,48.3\,,
\eea
i.e., the two-loop corrections shift the one-loop prediction 
away from the experimental value of 53.2. As already advertised in the 
Introduction, this set of numbers suggests a simple way to cure the problem:
It will be sufficient to introduce a significant enhancement of the Higgs $Z$ 
factors at low energies. This will be realized in the following using extra 
multiplets with extra Yukawa couplings. 

For completeness, we record the result which is obtained if the 
canonical (rather than the holomorphic) GUT threshold corrections are 
assumed to vanish. At the technical level, this corresponds simply
to replacing the `vector anomaly' contribution $-4.08$ above with
\be
\frac{24}{7} \, \ln \frac{\alpha_2(m_Z)}{\alpha_{\rm GUT}} \, - 3 \, \ln 
\frac{\alpha_3(m_Z)}{\alpha_{\rm GUT}} \, = \, - \,  0.67 \, - \, 3.17 \, = 
\, - \, 3.84\,.\label{cva}
\ee
The resulting prediction improves insignificantly. 

The smallness of this change is due to the weak dependence of the $\alpha_3$
prediction on the value of $\alpha_{\rm GUT}$ in Eq.~(\ref{cva}). This 
is the result of the approximate cancellation $24/7-3=3/7$. It is 
equivalent to the statement that the splitting between $m_{X,Y}$ and
$m_\Phi$ affects the $\alpha_3$ prediction only very weakly, which 
reflects the similarity of the ratios of the $b_i^\Phi$ and the ratios
of the corresponding MSSM coefficients $b_i$. 

Finally, we comment on the effect of MSSM Yukawa couplings which we
have so far neglected. At moderate $\tan\beta$, the top Yukawa coupling 
gives the dominant correction (cf. App.~B),
\be
\Delta_{\rm top}\left( \frac{2\pi}{\alpha_3} \right) = 0.32\,,
\ee
which is however far too small to cure the `two-loop' problem.

\section{Extra multiplets}

\subsection{Extra multiplets without Yukawa couplings}
\label{noyukawa}

To preserve one-loop gauge coupling unification in the most straightforward way
\footnote{See \cite{Calibbi:2009cp} for an alternative point of view}, we restrict 
our attention to complete SU(5) multiplets. More specifically, we focus on models 
with $n_5$ pairs of $\mathbf{5}+\overline{\mathbf{5}}$ and $n_{10}$ pairs of 
$\mathbf{10}+\overline{\mathbf{10}}$. At one loop, this leads to a modification, 
$b_i\to b_i'=b_i+n$, of the MSSM $\beta$ function coefficients, where 
$n = n_5 + 3 n_{10}$ is known as the `messenger index'. This, of course, does 
not affect the one-loop $\alpha_3$ prediction. 

The two-loop correction, given explicitly in Eq.~(\ref{da3}), changes only 
because of the modified $Z$ factors of the MSSM matter fields. We emphasize 
that Eq.~(\ref{da3}) does {\it not} need to be supplemented with $Z$ factors 
of the extra multiplets since these are assumed to decouple above $m_Z$.

The modified $Z$ factors are obtained from Eqs.~(\ref{zg}), as before, 
but now with the gauge group contributions (cf. Eq.~(\ref{zf}))
\bea
Z_{i}(m_Z) &=& \left(\frac{\widehat{\alpha}_{\rm GUT}}
{\alpha_i(m_Z)}\right)^{-\frac{2C_{F,i}}{b_i'}}  \,\, = \,\, 
\left(\,\frac{\frac{2\pi}{\alpha_i(m_Z)}}{ \frac{2\pi}{\alpha_{\rm GUT}} \, - 
\,n \ln \frac{M_{\rm GUT}}{m_Z}  }     \,\right)^{ - \frac{2C_{F,i}}{b_i + n} }\,.
\label{zmod} 
\eea
Here $\widehat{\alpha}_{\rm GUT}$ is the one-loop GUT-coupling in the 
presence of extra multiplets. Note that the one-loop GUT scale, 
$M_{\rm GUT}=2 \cdot 10^{16}$ GeV, is unaffected by the presence of the 
additional matter. For simplicity, we have at his stage neglected the 
necessary hierarchy between $m_Z$ and the scale at which the extra 
multiplets decouple.

The two-loop corrections related to the different matter field $Z$-factors
can be found in App.~C. Here we only list the overall two-loop 
prediction for $2\pi/\alpha_3$ for different  values of the $n$ parameter:
\begin{table}[ht]
\caption{Two-loop prediction for $2\pi/\alpha_3$} \label{twoloop}
\smallskip
\centering
\begin{tabular}{|c|c|c|c|}
\hline
\,\quad\quad\, $n$ \quad\quad\,\, & \,  Prediction for \,\,$2\pi/\alpha_3$\,\\
\hline
1    &      48.08   \\
\hline
2    &      47.91   \\
\hline
3    &      47.61   \\
\hline
4    &      46.81   \\
\hline
4.45 &      45.81   \\
\hline
\end{tabular}
\end{table}

The value $n=4.45$ formally corresponds to $\widehat{\alpha}_{\rm GUT}=1$.
Of course, this has to be interpreted as $n=5$ together with an 
appropriately raised decoupling scale. We see that with increasing $n$ 
the two-loop prediction for $\alpha_3$ becomes systematically worse. 

Of course, we could easily repeat the analysis using the assumption of
vanishing canonical GUT-thresholds. In this case, the increased 
value of the GUT-coupling enters the result also via the analogue of
Eq.~(\ref{cva}), with $\alpha_{\rm GUT}$ replaced by 
$\widehat{\alpha}_{\rm GUT}$. For $n=4.45$, this contribution changes 
from $-3.84$ to $-5.16$, giving $2\pi/\alpha_3=44.73$.

\subsection{Extra multiplets with Yukawa couplings - an 
 example with strong GUT coupling} \label{strong}

In the following we extend our analysis of Sect.~\ref{noyukawa} by allowing 
renormalizable couplings between the SU(5) matter and the MSSM
particles. To this end let us assume that there is at least one pair 
of $\mathbf{10}+\overline{\mathbf{10}}$ vector-like matter. In analogy 
to the MSSM, we denote its field content by $\mathbf{10}=(Q_e,U_e,E_e)$ 
and $\overline{\mathbf{10}}=(\overline{Q}_e,\overline{U}_e,\overline{E}_e)$, 
with an index `$e$' for `extra multiplet'. We can now introduce extra Yukawa 
couplings of the form:
\be
W \supset \, \kappa \, Q_e\, U_e H_U \, + \,  \overline{\kappa} \, 
\overline{Q}_e \overline{U}_e  H_D\,,\label{potential}
\ee
These new interactions modify the prediction for $\alpha_3$ solely
through their effect on the Higgs wavefunction renormalization 
factor $Z_H$ (where $Z_H$ stands for either $Z_{H_U}$ or $Z_{H_D}$).
We note that the couplings $\kappa$, $\overline{\kappa}$  can be 
extended to full $4\times4$ Yukawa matrices allowing for mixing between 
the three MSSM generations and the additional matter. However, these 
mixings have to be small due to FCNC constraints \cite{Graham:2009gy} 
(see also \cite{Chanowitz:2009mz,Bobrowski:2009ng}) and we neglect them.

In order to simplify the analysis in this section let us once again 
neglect any contributions arising from the MSSM Yukawa sector (based 
on our results from Sect.~\ref{prediction} and App.~B we expect those 
contributions to be small). This allows us to treat $\kappa$ and 
$\overline{\kappa}$ on  an equal footing. In particular we can set 
$\kappa = \overline{\kappa}$ and effectively deal with a single Yukawa 
coupling (say $\kappa$) and a single Higgs (calling this field $H$).

The one-loop RGE for $Z_H$ (with $t = \ln \mu$) reads
\be
2\pi\, \, \frac{d \ln Z_{H}}{dt} = - 3 \alpha_\kappa + 
\frac{3}{10}\alpha_1+ \frac{3}{2}\alpha_2\,,\label{fzrge} \,
\ee
where we have defined $\alpha_\kappa=\kappa_c^2/(4\pi)$, with $\kappa_c^2=
\kappa^2/(Z_H Z_{Q_e} Z_{U_e})$ the {\it canonical} Yukawa coupling. It is 
clear that a large $\kappa$ will drive $Z_H$ to larger values at the 
electroweak scale, improving the $\alpha_3$ prediction.

Equation~(\ref{fzrge}) entails the factorization property 
\be
Z_H = \underbrace{Z_H^G}_{\rm gauge \,\, part} \, \times 
\underbrace{Z_H^Y}_{\rm Yukawa \,\, part} \,, \label{fact}
\ee
where $Z_H^G$ and $Z_H^Y$ represent the contributions from the gauge 
couplings $\alpha_1$, $\alpha_2$ and the Yukawa coupling $\alpha_{\kappa}$. 
As before, the gauge part $Z^G_H$ is determined by Eqs.~(\ref{zg}) and 
(\ref{zmod}). We thus focus on the one-loop running of $Z_H^Y$:
\be
2\pi\, \, \frac{d \ln Z_{H}^Y}{dt} = - 3 \alpha_\kappa\,.
\label{zrge}
\ee
The corresponding one-loop RGE for $\alpha_{\kappa}$ reads
\be
2\pi \frac{d \ln \alpha_{\kappa}}{dt} \, = \, 6\,\alpha_{
\kappa} \, - \, \frac{16}{3} \alpha_3 - 3 \alpha_2 - 
\frac{13}{15}\alpha_1\,.\label{krge}
\ee
In the following we will analyze a model in which the value of the low-scale
$Z_{H}^Y$-factor can be obtained in a completely analytical manner. To this 
end let us assume that both the extra Yukawa couplings as well as the gauge 
couplings begin their evolution at the strong-coupling point at the high 
scale:
\be
\widehat{\alpha}_{GUT}\sim\alpha_\kappa(M_{\rm GUT})\sim1\,.\label{sy}
\ee
Formally this corresponds to $n=4.45$. In this case the relations 
\be
\alpha_2 \, = \, \frac{b_3'}{b_2'}\,\alpha_3  \quad\quad\quad\quad\quad  
\alpha_1 \, = \, \frac{b_3'}{b_1'}\,\alpha_3\label{aa}
\ee
will be approximately valid at all energies significantly below the GUT 
scale.\footnote{
One 
writes $\,\alpha_i^{-1}(\mu)={\cal O}(1)+b_i'\ln
\left(\frac{M_{\rm GUT}}{\mu}\right)\,$ and neglects the ${\cal O}(1)$ term.
}
Equation~(\ref{krge}) now takes the form 
\be
2\pi \frac{d \ln \alpha_{\kappa}}{dt} = 
6\alpha_{\kappa} - \alpha_3 \left( \frac{16}{3} + 
\frac{3 b_3'}{b_2'} +\frac{13b_3'}{15b_1'}\right)\,.
\ee
Using the one-loop RGEs for the gauge couplings
\be
\frac{d\ln\alpha_i^{-1}}{dt}\,=-b_i'\,\frac{\alpha_i}{2\pi}\,,
\ee
this can be further rewritten as
\be
2\pi \frac{d \ln(\alpha_{\kappa}/\alpha_3)}{dt} = 
6\alpha_{\kappa} - \alpha_3 \left( \frac{16}{3} + 
\frac{3 b_3'}{b_2'} +\frac{13b_3'}{15b_1'} + b_3'\right)\,.\label{rrge}
\ee
Eq.~(\ref{rrge}) has an infrared-stable fixed point of the 
Pendleton-Ross type \cite{Pendleton:1980as} given by
\be
\alpha_{\kappa} = 1.37 \alpha_3\qquad\mbox{for}\qquad n=4.45\,.
\label{ka}
\ee
As a result of fast initial evolution, the ratios of Yukawa and gauge 
couplings quickly reach the fixed-point regime, which is then maintained 
all the way down to the weak scale. From Eq.~(\ref{zrge}) we now have
\be
\frac{d \ln Z_H^Y}{dt} = - 
3\cdot 1.37 \, \frac{\alpha_3}{2\pi} = \frac{4.11}{b_3'}
\frac{d\ln\alpha_3^{-1}}{dt}
\ee
and
\be
\ln Z_H^Y(m_Z) = 2.83 \ln (\alpha_{\rm GUT}/\alpha_3(m_Z)) = 6.05\,.
\ee
The resulting correction to the $\alpha_3$ prediction is
\be
\Delta_{\kappa}\left(\frac{2\pi}{\alpha_3}\right) = 
\frac{9}{7}\,\ln Z^Y_H(m_Z) = 7.78\,,
\ee
which is just sufficient to compensate the negative two-loop effects in
the last line of Table~\ref{twoloop}. Of course, at this stage our 
promising results should be taken with caution. Specifically, when talking 
about a strongly-coupled unified theory, one faces the danger of potentially 
large and incalculable corrections at the GUT scale which could, in 
principle, render the entire two-loop analysis obsolete. We postpone the
discussion of these issues to Sect.~\ref{thresh}. We also note that the 
influence of extra Yukawas \textit{above} the GUT scale on the unified coupling
has recently been discussed in \cite{Aranda:2009wh}.

\subsection{Effect of extra Yukawas in models with 
 perturbative gauge couplings}\label{perturbative}

In this section we extend our previous results to a more general setting
by lifting some of the simplifying ad hoc assumptions that were made so far.
This means in particular that we introduce an explicit decoupling scale $M$
for the extra matter fields. Also, from now on the messenger index $n$ is 
allowed to attain only integer values (however, we only allow for $n\ge3$ 
since we assume at least one $\mathbf{10}+\overline{\mathbf{10}}$ pair). The 
superpotential in the extended Yukawa sector is still specified by 
Eq.~(\ref{potential}). In order to improve the accuracy of our predictions 
we will also take into account the contributions from the MSSM top Yukawa 
coupling. Since this coupling enters the one-loop RGEs of $\alpha_{\kappa}$ 
and $\alpha_{\overline{\kappa}}$ in a different manner, this step explicitly 
breaks the symmetry of our model with respect to $\kappa$ and $\overline{\kappa}$. 
In particular, from now on we will distinguish explicitly between these two couplings.

The most important constraint which we impose on the models in this section 
is perturbativity of the gauge couplings. Among other things this implies 
that the analytical approach we developed in the previous section is no longer 
applicable. Instead we will employ an alternative technique, which will allow us 
to handle our models semi-analytically.

To this end we will have to deal with the two different regimes of the theory 
(the high-energy regime above $M$ and the low-energy regime below $M$) 
separately. Let us focus on the high-energy theory first. As was mentioned before,
any MSSM field obeys a factorization property analogous to Eq.~(\ref{fact}). This 
means that the gauge and Yukawa contributions to the matter field $Z$-factors 
decouple and can be analyzed independently. The gauge parts of the $Z$-factors are 
obtained from Eq.~(\ref{zmod}) by simply replacing $m_Z \to M$. In order to get 
the Yukawa parts we will need the one-loop RGEs for the $Z^{Y}$ (above $M$):
\bea
2\pi\, \, \frac{d \ln Z_{H_U}^Y}{dt} &=& - 3 \alpha_t - 3 \alpha_{\kappa} \,,
\qquad 2\pi\, \, \frac{d \ln Z_{H_D}^Y}{dt} =  - 3 \alpha_{\overline{\kappa}} 
\,,\nonumber \\
2\pi\, \, \frac{d \ln Z_{U_3}^Y}{dt} &=& - 2 \alpha_t \,, \qquad\qquad \quad
2\pi\, \, \frac{d \ln Z_{Q_3}^Y}{dt} = -   \alpha_t \,.\label{zfac}
\eea
Since we regard only the $\kappa$, $\overline{\kappa}$ and $y_t$ Yukawas as
non-vanishing, all other $Z^Y$-factors are irrelevant for our analysis. In the 
following we will also need the one-loop equations for the three aforementioned 
Yukawa couplings:
\bea
2\pi \,\frac{d \ln \alpha_t}{dt} \, &=& \,6\alpha_t + 3\,\alpha_{\kappa} - 
\frac{13}{15}\,\alpha_1  \, - \, 3\alpha_2  \, - \,\frac{16}{3}\,\alpha_3\,  
\nonumber \\
2\pi \,\frac{d \ln \alpha_{\kappa}}{dt} \, &=& \, 6\alpha_{\kappa} + 
3  \alpha_t - \frac{13}{15}\,\alpha_1 \, - \, 3\alpha_2 \, - \,
\frac{16}{3}\,\alpha_3\,  \label{yuk} \\
2\pi \,\frac{d \ln \alpha_{\overline{\kappa}}}{dt} \, &=& \, 6
\alpha_{\overline{\kappa}} - \frac{13}{15}\,\alpha_1 \,  - \, 3\alpha_2  
\, - \,\frac{16}{3}\,\alpha_3\,  \nonumber
\eea
Note that these identities generalize Eq.~(\ref{krge}) to the case of non-vanishing
$\alpha_t$. Following \cite{AmelinoCamelia:1998tm} we can combine Eqs.~(\ref{yuk}) 
and (\ref{zfac}) to get a closed analytic expression for the Yukawa parts of the 
relevant $Z$-factors:
\bea
Z^Y_{Q_3}(M)&=&\left( \frac{\widehat{\alpha}_{\rm GUT}}{\alpha_1(M)} 
\right)^{\frac{1}{9} \,\frac{13}{15 \, b_1'}} \left( \frac{\widehat{\alpha}_{\rm 
GUT}}{\alpha_2(M)} \right)^{\frac{1}{9}\,\frac{3}{b_2'}} \left( \frac{\widehat
{\alpha}_{\rm GUT}}{\alpha_3(M)} \right)^{\frac{1}{9}\, \frac{16}{3 \, b_3'}} 
\left(\frac{\alpha_{t,\rm GUT}}{\alpha_t(M)} \right)^{\frac{2}{9}} \left(\frac
{\alpha_{\kappa, \rm GUT}}{\alpha_{\kappa}(M)}\right)^{-\frac{1}{9}} \nonumber \\
Z^Y_{U_3}(M)&=&\left( \frac{\widehat{\alpha}_{\rm GUT}}{\alpha_1(M)} \right)
^{\frac{2}{9} \,\frac{13}{15 \, b_1'}} \left( \frac{\widehat{\alpha}_{\rm GUT}}
{\alpha_2(M)} \right)^{\frac{2}{9}\,\frac{3}{b_2'}} \left( \frac{\widehat{\alpha}
_{\rm GUT}}{\alpha_3(M)} \right)^{\frac{2}{9}\,\frac{16}{3\,b_3'}} \left(\frac
{\alpha_{t,\rm GUT}}{\alpha_t(M)} \right)^{\frac{4}{9}}\left(\frac{\alpha_{\kappa,
\rm GUT}}{\alpha_{\kappa}(M)}\right)^{-\frac{2}{9}} \nonumber \\
Z^Y_{H_D}(M)&=&\left( \frac{\widehat{\alpha}_{\rm GUT}}{\alpha_1(M)} \right)^{\frac{1}{2} 
\,\frac{13}{15 \, b_1'}} \left( \frac{\widehat{\alpha}_{\rm GUT}}{\alpha_2(M)} \right)
^{\frac{1}{2}\,\frac{3}{b_2'}} \left( \frac{\widehat{\alpha}_{\rm GUT}}{\alpha_3(M)} 
\right)^{\frac{1}{2}\, \frac{16}{3 \, b_3'}} \left(\frac{\alpha_{\overline{\kappa},\rm GUT}}
{\alpha_{\overline{\kappa}}(M)} \right)^{\frac{1}{2}} \quad\quad\quad\quad\quad \label{bigg}\\
Z^Y_{H_U}(M)&=&\left( \frac{\widehat{\alpha}_{\rm GUT}}{\alpha_1(M)} \right)^{\frac{2}{3} \,
\frac{13}{15 \, b_1'}} \left( \frac{\widehat{\alpha}_{\rm GUT}}{\alpha_2(M)} \right)
^{\frac{2}{3}\,\frac{3}{b_2'}} \left( \frac{\widehat{\alpha}_{\rm GUT}}{\alpha_3(M)} \right)
^{\frac{2}{3}\, \frac{16}{3 \, b_3'}} \left( \frac{\alpha_{t,\rm GUT}}{\alpha_t(M)} \right)^{
\frac{1}{3}} \left(\frac{\alpha_{\kappa,\rm GUT}}{\alpha_{\kappa}(M)}\right)^{
\frac{1}{3}} \,.\nonumber 
\eea

The theory below the scale $M$ is the MSSM. Therefore we can apply our 
formulas from Sect.~\ref{prediction} and App.~B. The only modification 
is that we now integrate from $m_Z$ to $M$ rather than to $M_{\rm GUT}$. As 
an illustrative example we record the result for the Yukawa part 
of the $Z_{H_U}$ factor:
\be
\frac{Z^Y_{H_U}(m_Z)}{Z^Y_{H_U}(M)} = \left(\,\frac{\alpha_1(M)}
{\alpha_1(m_Z)} \right)^{\frac{13}{30\,b_1}} 
\left(\,\frac{\alpha_2(M)}{\alpha_2(m_Z)}
\right)^{\frac{3}{2\,b_2}}
\left(\,\frac{\alpha_3(M)}{\alpha_3(m_Z)}\,
\right)^{\frac{8}{3\,b_3}}
\left(\,\frac{\alpha_t(M)}{\alpha_t(m_Z)}\right)^{\frac{1}
{2}}\, \label{num}
\ee
The calculation of the other $Z$-factors proceeds in a similar manner. 
The low-scale values $Z(m_Z)$ are then obtained by multiplying the
expressions for $Z(M)$ and $Z(m_Z)/Z(M)$.

The brackets in Eqs.~(\ref{bigg}) and (\ref{num}) involving gauge couplings
can be evaluated analytically (by using the respective one-loop values).
To calculate the Yukawa brackets we have  solved the one-loop RGEs 
in Eq.~(\ref{yuk}) numerically by evolving them from the GUT down to the 
electroweak scale and using a $\theta$-function approximation at the 
decoupling scale $M$. The resulting low-scale values for the three
Yukawa couplings were then substituted in Eqs.~(\ref{bigg}) and 
(\ref{num}). In Tables \ref{n4} and \ref{n5} we have listed the two-loop 
corrections to $2\pi/\alpha_3$ for different values of the two parameters
$M$ and $n$. We have also tested the sensitivity of our 
results against variations of the initial values $\,\alpha_{\kappa}(M_{\rm GUT})
\,$ and  $\,\alpha_{\overline{\kappa}}(M_{\rm GUT})\,$ (we remark that the values 
$\alpha_{\kappa}(M_{\rm GUT})=\alpha_{\overline{\kappa}}(M_{\rm GUT})=0.228, 
0.457, 0.913$ correspond to $1/6,1/3,2/3$ times the fixed point value $1.37$ 
of the extra Yukawa couplings). In each case the initial value for the top 
Yukawa coupling at $M_{\rm GUT}$ has been adjusted to reproduce the correct 
low-energy  parameter $y_t(m_Z) = 0.99$ (see also App.~B).

A quick glance at Tables \ref{n4} and \ref{n5} reveals that models with
$n=4$ are favoured over their $n=5$ counterparts. The former are also 
of particular interest because of the low value of $M$ \cite{Babu:2008ge,
Graham:2009gy,Martin:2010dc,Jiang:2006hf}. We have intentionally
omitted the $n=3$ case because the $n=3$ models are unable to generate a 
sufficiently large GUT coupling (say $\alpha_{\rm GUT} \ge 0.2$).

\begin{table}[ht] 
\caption{Numerical results $n=4$ (one pair of extra Yukawa couplings)} \label{n4}
\smallskip
\centering
\begin{tabular}{|l|c|c|c|c|c|c|c|}
\hline
$M$(GeV) & $\alpha_{\rm GUT}$ & 
$\alpha_{\kappa, \rm GUT}=\alpha_{\overline{\kappa}, \rm GUT}$ &
$\alpha_{\kappa}(M)$ & $\alpha_{\overline{\kappa}}(M)$ &
$y_t(M_{\rm GUT})$ & $2\pi/\alpha_3$   \\
\hline
\bf{500} & \bf{0.227} & \bf{0.228} & \bf{0.102} & \bf{0.134} & \bf{0.68} & \bf{51.10}   \\
\hline
\textcolor{gray}{500} & \textcolor{gray}{0.227} & \bf{0.457} & \bf{0.102} 
& \bf{0.134} & \bf{0.84} & \bf{51.55}   \\
\hline
\textcolor{gray}{500} & \textcolor{gray}{0.227} & \bf{0.913} & \bf{0.103} 
& \bf{0.134} & \bf{1.01} & \bf{51.99}   \\
\hline
\textcolor{gray}{500} & \textcolor{gray}{0.227} & \bf{2.000} & \bf{0.103} & 
\bf{0.134} & \bf{1.26} & \bf{52.50}   \\
\hline
\textcolor{gray}{500} & \textcolor{gray}{0.227} & \bf{4.000} & \bf{0.103} 
& \bf{0.134} & \bf{1.52} & \bf{52.95}   \\
\hline
\textcolor{gray}{500} & \textcolor{gray}{0.227} & \bf{6.000} & \bf{0.103} 
& \bf{0.134} & \bf{1.70} & \bf{53.21}   \\
\hline
\bf{300} & \bf{0.245} & \bf{0.457} & \bf{0.105} & \bf{0.137} & \bf{0.78} & \bf{51.67}   \\
\hline
\textcolor{gray}{300} & \textcolor{gray}{0.245} & \bf{2.000} & \bf{0.105} 
& \bf{0.137} & \bf{1.18} & \bf{52.62}   \\
\hline
\textcolor{gray}{300} & \textcolor{gray}{0.245} & \bf{4.000} & \bf{0.105} 
& \bf{0.137} & \bf{1.42} & \bf{53.07}   \\
\hline
\textcolor{gray}{300} & \textcolor{gray}{0.245} & \bf{6.000} & \bf{0.105} 
& \bf{0.137} & \bf{1.58} & \bf{53.33}   \\
\hline
\end{tabular}
\end{table}

\medskip

\begin{table}[ht]
\caption{Numerical results $n=5$ (one pair of extra Yukawa couplings)} \label{n5}
\smallskip
\centering
\begin{tabular}{|l|c|c|c|c|c|c|c|}
\hline
$M$(GeV) & $\alpha_{\rm GUT}$ & 
$\alpha_{\kappa, \rm GUT}=\alpha_{\overline{\kappa}, \rm GUT}$ &
$\alpha_{\kappa}(M)$ & $\alpha_{\overline{\kappa}}(M)$ &
$y_t(M_{\rm GUT})$ & $2\pi/\alpha_3$   \\
\hline
$\mathbf{25 \cdot 10^4}$ & \bf{0.229} & \bf{0.228} & \bf{0.097} & \bf{0.121} & 
\bf{0.70} & \bf{50.34}   \\
\hline
\textcolor{gray}{$25 \cdot 10^4$}  & \textcolor{gray}{0.229} & \bf{0.457} & 
\bf{0.097} & \bf{0.122} & \bf{0.86} & \bf{50.79}   \\
\hline
\textcolor{gray}{$25 \cdot 10^4$}  & \textcolor{gray}{0.229} & \bf{0.913} & 
\bf{0.098} & \bf{0.122} & \bf{1.04} & \bf{51.24}   \\
\hline
\textcolor{gray}{$25 \cdot 10^4$} & \textcolor{gray}{0.229} & \bf{2.000} & 
\bf{0.098} & \bf{0.122} & \bf{1.29} & \bf{51.74}   \\
\hline
\textcolor{gray}{$25 \cdot 10^4$}  & \textcolor{gray}{0.229} & \bf{4.000} & 
\bf{0.098} & \bf{0.122} & \bf{1.56} & \bf{52.19}   \\
\hline
\textcolor{gray}{$25 \cdot 10^4$}  & \textcolor{gray}{0.229} & \bf{6.000} & 
\bf{0.098} & \bf{0.122} & \bf{1.73} & \bf{52.45}   \\
\hline
\end{tabular}
\end{table}

It is important to note that the low-energy couplings $\alpha_{\kappa}(m_Z)$ 
and $\alpha_{\overline{\kappa}}(m_Z)$ are virtually insensitive to their 
input values at the GUT scale. This observation indicates a very straightforward 
way of increasing the two-loop prediction for $2\pi/\alpha_3$ -- namely by 
taking the input parameters $\,\alpha_{\kappa, \rm GUT}\,$ and $\,\alpha_{
\overline{\kappa}, \rm GUT}$ as large as possible. Note also that (in 
contradistinction to gauge couplings) the Yukawas do not exhibit any flavor 
enhancement (cf. Sect.~\ref{thresh}). Therefore the actual expansion parameters 
are $\alpha_{\kappa}/4\pi$ and $\alpha_{\overline{\kappa}}/4\pi$, which means 
that the strong coupling condition reads $\alpha_{\kappa}/4\pi \sim 
\alpha_{\overline{\kappa}}/4\pi \sim 1$. Following this line of thought we have 
considered models whose input values $\alpha_{\kappa,\rm GUT}$ and 
$\alpha_{\overline{\kappa},\rm GUT}$ are as high as 6.0. 

We once again emphasize that the gauge couplings in this section are only allowed
to attain perturbative values (in contradistinction to their Yukawa counterparts). 
As will be argued in Sect.~\ref{thresh} the lowering of the unified coupling from 
$\,\alpha_{\rm GUT} \sim \mathcal{O}(1)\,$ down to $\,\sim 0.2\,$ can potentially 
have a dramatic impact on the magnitude and calculability of the high-scale threshold 
corrections around the GUT scale.

\subsection{Models with further Yukawa couplings}\label{extended}

In the following we will investigate the principal effect of introducing 
further non-standard Yukawa couplings. We focus on two extensions: First we 
consider adding new $\mathbf{10}+\overline{\mathbf{10}}$ pairs which couple 
to the observable sector through an interaction analogous to Eq.(\ref{potential}).
In particular we increase the messenger index to $n\ge6$. Second, we explore a 
different possibility by restricting ourselves to the $n=4$ case but allowing 
for renormalizable interactions between the $\mathbf{10}+\overline{\mathbf{10}}$ 
and the $\mathbf{5}+\overline{\mathbf{5}}$ fields.

Following this line of thought let us now increase the number of additional 
vector-like multiplets to $n=6$ by introducing two $\mathbf{10}+
\overline{\mathbf{10}}$ pairs. We postulate a superpotential of the form:
\be
W \supset \, \kappa \, Q_e\, U_e H_U \, + \,  \overline{\kappa} \, 
\overline{Q}_e \overline{U}_e  H_D\, + \kappa' \, Q_e'\, U_e' \,H_U \,
+ \,  \overline{\kappa}' \, \overline{Q}_e' \overline{U}_e'  \,H_D\,
\label{extpotential}
\ee
where the primed fields denote the matter content of the new
$\mathbf{10}+\overline{\mathbf{10}}$ pair. In the following we will 
treat the two Yukawa pairs $(\kappa,\overline{\kappa})$ and $(\kappa',
\overline{\kappa}')$ on an equal footing, i.e. we assume that 
$\kappa = \kappa'$ and $\overline{\kappa} = \overline{\kappa}'$.
Note that by going from a superpotential of the form (\ref{potential})
to superpotential of the form (\ref{extpotential}) we only change
the running of the \textit{Yukawa} part of the $Z$-factors \textit{above} 
the decoupling scale. The relevant one-loop RGEs are
\bea
2\pi\, \, \frac{d \ln Z_{H_U}^Y}{dt} &=& - 3 \alpha_t - 6 \alpha_{\kappa} \,,
\qquad 2\pi\, \, \frac{d \ln Z_{H_D}^Y}{dt} =  - 6 \alpha_{\overline{\kappa}} 
\,,\nonumber \\
2\pi\, \, \frac{d \ln Z_{U_3}^Y}{dt} &=& - 2 \alpha_t \,, \qquad\qquad\,\,\,\,
\,\,2\pi\, \, \frac{d \ln Z_{Q_3}^Y}{dt} = -   \alpha_t \label{zfmod}
\eea
for the $Z^Y$ and
\bea
2\pi \,\frac{d \ln \alpha_t}{dt} \, &=& \,6\alpha_t + 6\,\alpha_{\kappa} - 
\frac{13}{15}\,\alpha_1  \, - \, 3\alpha_2  \, - \,\frac{16}{3}\,\alpha_3\,  
\nonumber \\
2\pi \,\frac{d \ln \alpha_{\kappa}}{dt} \, &=& \, 9\alpha_{\kappa} + 
3  \alpha_t - \frac{13}{15}\,\alpha_1 \, - \, 3\alpha_2 \, - \,
\frac{16}{3}\,\alpha_3\,  \label{yukmod} \\
2\pi \,\frac{d \ln \alpha_{\overline{\kappa}}}{dt} \, &=& \, 9
\alpha_{\overline{\kappa}} - \frac{13}{15}\,\alpha_1 \,  - \, 3\alpha_2  
\, - \,\frac{16}{3}\,\alpha_3\,  \nonumber
\eea
for the Yukawas.

Combining Eqs.~(\ref{zfmod}) and (\ref{yukmod}) we arrive at the analogue
of Eq.~(\ref{bigg}):
\bea
Z^Y_{Q_3}(M)&\!\!=\!\!&\left( \frac{\widehat{\alpha}_{\rm GUT}}{\alpha_1(M)} \right)
^{\frac{1}{12} \,\frac{13}{15 \, b_1'}} \left( \frac{\widehat{\alpha}_{\rm GUT}}
{\alpha_2(M)} \right)^{\frac{1}{12}\,\frac{3}{b_2'}} \left( \frac{\widehat{\alpha}_
{\rm GUT}}{\alpha_3(M)} \right)^{\frac{1}{12}\,\frac{16}{3 \, b_3'}} \left( \frac{
\alpha_{t,\rm GUT}}{\alpha_t(M)} \right)^{\frac{1}{4}} \left(\frac{\alpha_{\kappa,
\rm GUT}}{\alpha_{\kappa}(M)}\right)^{-\frac{1}{6}} \nonumber \\
Z^Y_{U_3}(M)&\!\!=\!\!&\left( \frac{\widehat{\alpha}_{\rm GUT}}{\alpha_1(M)} \right)
^{\frac{1}{6} \,\frac{13}{15 \, b_1'}} \left( \frac{\widehat{\alpha}_{\rm GUT}}{
\alpha_2(M)} \right)^{\frac{1}{6}\,\frac{3}{b_2'}} \left( \frac{\widehat{\alpha}_{\rm GUT}}
{\alpha_3(M)} \right)^{\frac{1}{6}\,\frac{16}{3 \, b_3'}} \left( \frac{\alpha_{t,\rm GUT}}
{\alpha_t(M)} \right)^{\frac{1}{2}}  \left(\frac{\alpha_{\kappa,\rm GUT}}{\alpha_{\kappa}
(M)}\right)^{-\frac{1}{3}} \nonumber \\
Z^Y_{H_D}(M)&\!\!=\!\!&\left( \frac{\widehat{\alpha}_{\rm GUT}}{\alpha_1(M)} \right)
^{\frac{2}{3} \,\frac{13}{15 \, b_1'}} \left( \frac{\widehat{\alpha}_{\rm GUT}}{\alpha_2(M)}
 \right)^{\frac{2}{3}\,\frac{3}{b_2'}} \left( \frac{\widehat{\alpha}_{\rm GUT}}{\alpha_3(M)} 
\right)^{\frac{2}{3}\,\frac{16}{3 \, b_3'}} \left(\frac{\alpha_{\overline{\kappa},\rm GUT}}
{\alpha_{\overline{\kappa}}(M)} \right)^{\frac{2}{3}} \label{biggmod}\\
Z^Y_{H_U}(M)&\!\!=\!\!&\left( \frac{\widehat{\alpha}_{\rm GUT}}{\alpha_1(M)} \right)
^{\frac{3}{4} \,\frac{13}{15 \, b_1'}} \left( \frac{\widehat{\alpha}_{\rm GUT}}{\alpha_2(M)} 
\right)^{\frac{3}{4}\,\frac{3}{b_2'}} \left( \frac{\widehat{\alpha}_{\rm GUT}}{\alpha_3(M)} 
\right)^{\frac{3}{4}\, \frac{16}{3 \, b_3'}} \left( \frac{\alpha_{t,\rm GUT}}{\alpha_t(M)} 
\right)^{\frac{1}{4}} \, \left(\frac{\alpha_{\kappa,\rm GUT}}{\alpha_{\kappa}(M)}\right)^{
\frac{1}{2}} \,.\nonumber
\eea

The results from our numerical analysis are listed in Table \ref{n=6}. The prediction 
for $\alpha_3$ improves only slightly in comparison to the `optimal' $n=4$ models with 
a single pair of extra Yukawas.

\begin{table}[ht]
\caption{Numerical results n=6 (two pairs of extra Yukawa couplings)}
\smallskip
\label{n=6}
\centering
\begin{tabular}{|l|c|c|c|c|c|c|c|}
\hline
$M$(GeV) & $\alpha_{\rm GUT}$ & 
$\alpha_{\kappa, \rm GUT}=\alpha_{\overline{\kappa}, \rm GUT}$ &
$\alpha_{\kappa}(M)$ & $\alpha_{\overline{\kappa}}(M)$ &
$y_t(M_{\rm GUT})$ & $2\pi/\alpha_3$   \\
\hline
$\mathbf{17 \cdot 10^6}$ & \bf{0.227} & \bf{0.228} & \bf{0.064} & \bf{0.080} 
& \bf{1.26} & \bf{51.08}   \\
\hline
\textcolor{gray}{$17 \cdot 10^6$} & \textcolor{gray}{0.227} & \bf{0.457} & 
\bf{0.065} & \bf{0.080} & \bf{1.68} & \bf{51.66}   \\
\hline
\textcolor{gray}{$17 \cdot 10^6$} & \textcolor{gray}{0.227} & \bf{0.913} & 
\bf{0.065} & \bf{0.081} & \bf{2.24} & \bf{52.25}   \\
\hline
\textcolor{gray}{$17 \cdot 10^6$} & \textcolor{gray}{0.227} & \bf{2.000} & 
\bf{0.065} & \bf{0.081} & \bf{3.08} & \bf{52.92}   \\
\hline
\textcolor{gray}{$17 \cdot 10^6$} & \textcolor{gray}{0.227} & \bf{4.000} & 
\bf{0.065} & \bf{0.081} & \bf{4.06} & \bf{53.52} \\
\hline
\textcolor{gray}{$17 \cdot 10^6$} & \textcolor{gray}{0.227} & \bf{6.000} & 
\bf{0.065} & \bf{0.081} & \bf{4.77} & \bf{53.86}   \\
\hline
\end{tabular}
\end{table}

The next type of models we consider contain one pair of $\mathbf{10}+
\overline{\mathbf{10}}$ and one pair of $\mathbf{5}+\overline{\mathbf{5}}$ 
extra multiplets. Using the standard decomposition $\mathbf{5} = (\overline{D}_e,
\overline{L}_e)$ and $\overline{\mathbf{5}} = (D_e,L_e)$ we introduce a superpotential 
of the form:
\be
W \supset \, \kappa \, Q_e\, U_e H_U \, + \,  \overline{\kappa} \, 
\overline{Q}_e \,\overline{U}_e  H_D\, + \, \lambda \, Q_e \, D_e H_D \,
+ \, \overline{\lambda} \, \overline{Q}_e \, \overline{D}_e H_U \label{n4potential}
\ee
Note that this is a direct extension of the $n=4$ models from Sect.~\ref{perturbative} 
-- we have simply added two new interactions to our superpotential. The 
calculation of the two-loop $\alpha_3$-correction proceeds exactly as before. 
Here we only list the relevant one-loop RGEs above the decoupling scale $M$:
\bea
2\pi\, \, \frac{d \ln Z_{H_U}^Y}{dt} &=& - 3 \alpha_t - 3 \alpha_{\kappa} 
-3 \alpha_{\overline{\lambda}} \,\,,\qquad 2\pi\, \, \frac{d \ln Z_{H_D}^Y}{dt} =  
- 3 \alpha_{\overline{\kappa}} -3 \alpha_{\lambda} \,,\nonumber \\
2\pi\, \, \frac{d \ln Z_{U_3}^Y}{dt} &=& - 2 \alpha_t \,, \qquad\qquad\qquad\qquad
2\pi\, \, \frac{d \ln Z_{Q_3}^Y}{dt} = -   \alpha_t \label{zfmodmod}
\eea
for the matter field $Z^Y$-factors and
\bea
2\pi \,\frac{d \ln \alpha_t}{dt} \, &=& \,6\alpha_t + 3\,\alpha_{\kappa} +
3\,\alpha_{\overline{\lambda}} \, - \,\frac{13}{15}\,\alpha_1  \, - \, 3\alpha_2
\, - \,\frac{16}{3}\,\alpha_3\, \nonumber \\
2\pi \,\frac{d \ln \alpha_{\kappa}}{dt} \, &=& \, 6\alpha_{\kappa} + 3\,\alpha_{t}
+3\,\alpha_{\overline{\lambda}} - \frac{13}{15}\,\alpha_1 \, - \, 3\alpha_2 \, - \,
\frac{16}{3}\,\alpha_3\,  \nonumber \\
2\pi \,\frac{d \ln \alpha_{\overline{\kappa}}}{dt} \, &=& \, 6 \alpha_{\overline{
\kappa}} + 3\, \alpha_{\lambda} - \frac{13}{15}\,\alpha_1 \,  - \, 3\alpha_2  
\, - \,\frac{16}{3}\,\alpha_3\,  \label{yukn4} \\
2\pi \,\frac{d \ln \alpha_{\lambda}}{dt} \, &=& \, 6 \alpha_{\lambda} + 3\, 
\alpha_{\overline{\kappa}} - \frac{7}{15}\,\alpha_1 \,  - \, 3\alpha_2  
\, - \,\frac{16}{3}\,\alpha_3\, \nonumber \\
2\pi \,\frac{d \ln \alpha_{\overline{\lambda}}}{dt} \, &=& \, 6\alpha_{
\overline{\lambda}} + 3\,\alpha_{t} + 3\,\alpha_{\kappa} - \frac{7}{15}\,\alpha_1 
\, - \, 3\alpha_2 \, - \,\frac{16}{3}\,\alpha_3\, \nonumber
\eea
for the Yukawa couplings. As before we have defined $\,\alpha_{\lambda} = 
\lambda^2/(4\pi)\,$ and $\,\alpha_{\overline{\lambda}} = \overline{\lambda}^2/(4\pi)\,$.
The modification of Eq.(\ref{bigg}) reads
\bea
Z^Y_{Q_3}(M)&\!\!=\!\!&\left( \frac{\widehat{\alpha}_{\rm GUT}}{\alpha_1(M)} \right)
^{\frac{1}{12} \,\frac{19}{15 \, b_1'}} \left( \frac{\widehat{\alpha}_{\rm GUT}}
{\alpha_2(M)} \right)^{\frac{1}{12}\,\frac{3}{b_2'}} \left( \frac{\widehat{\alpha}_
{\rm GUT}}{\alpha_3(M)} \right)^{\frac{1}{12}\,\frac{16}{3 \, b_3'}} \left( \frac{
\alpha_{t,\rm GUT}}{\alpha_t(M)} \right)^{\frac{1}{4}} \,\times \nonumber\\
&&\qquad\qquad\qquad\qquad\qquad\qquad\qquad\,\,\,\times \, \left(\frac{\alpha_{\kappa,\rm
GUT}}{\alpha_{\kappa}(M)}\right)^{-\frac{1}{12}} \left(\frac{\alpha_{\overline{\lambda},
\rm GUT}}{\alpha_{\overline{\lambda}}(M)}\right)^{-\frac{1}{12}}  \label{bigmod} \\
\noalign{\smallskip}
Z^Y_{U_3}(M)&\!\!=\!\!&\left( \frac{\widehat{\alpha}_{\rm GUT}}{\alpha_1(M)} \right)
^{\frac{1}{6} \,\frac{19}{15 \, b_1'}} \left( \frac{\widehat{\alpha}_{\rm GUT}}{
\alpha_2(M)} \right)^{\frac{1}{6}\,\frac{3}{b_2'}} \left( \frac{\widehat{\alpha}_{\rm GUT}}
{\alpha_3(M)} \right)^{\frac{1}{6}\,\frac{16}{3 \, b_3'}} \left( \frac{\alpha_{t,\rm GUT}}
{\alpha_t(M)} \right)^{\frac{1}{2}} \,\times \, \nonumber\\
&&\qquad\qquad\qquad\qquad\qquad\qquad\qquad\,\,\,\times\, \left(\frac{\alpha_{\kappa,\rm 
GUT}}{\alpha_{\kappa}(M)}\right)^{-\frac{1}{6}}  \left(\frac{\alpha_{\overline{\lambda},
\rm GUT}}{\alpha_{\overline{\lambda}}(M)}\right)^{-\frac{1}{6}}  \,\,.\nonumber
\eea

\bea
Z^Y_{H_D}(M)&\!\!=\!\!&\left( \frac{\widehat{\alpha}_{\rm GUT}}{\alpha_1(M)} \right)
^{\frac{1}{3} \,\frac{20}{15 \, b_1'}} \left( \frac{\widehat{\alpha}_{\rm GUT}}{\alpha_2(M)}
 \right)^{\frac{2}{3}\,\frac{3}{b_2'}} \left( \frac{\widehat{\alpha}_{\rm GUT}}{\alpha_3(M)} 
\right)^{\frac{2}{3}\,\frac{16}{3 \, b_3'}} \left(\frac{\alpha_{\overline{\kappa},\rm GUT}}
{\alpha_{\overline{\kappa}}(M)} \right)^{\frac{1}{3}}  \left(\frac{\alpha_{\lambda,
\rm GUT}}{\alpha_{\lambda}(M)} \right)^{\frac{1}{3}}  \nonumber \\
Z^Y_{H_U}(M)&\!\!=\!\!&\left( \frac{\widehat{\alpha}_{\rm GUT}}{\alpha_1(M)} \right)
^{\frac{1}{4} \,\frac{33}{15 \, b_1'}} \left( \frac{\widehat{\alpha}_{\rm GUT}}{\alpha_2(M)} 
\right)^{\frac{3}{4}\,\frac{3}{b_2'}} \left( \frac{\widehat{\alpha}_{\rm GUT}}{\alpha_3(M)} 
\right)^{\frac{3}{4}\, \frac{16}{3 \, b_3'}} \left( \frac{\alpha_{t,\rm GUT}}{\alpha_t(M)} 
\right)^{\frac{1}{4}} \,\times \nonumber\\
&&\qquad\qquad\qquad\qquad\qquad\qquad\qquad\qquad\qquad\,\,\,\,\,\times\, 
\left(\frac{\alpha_{\kappa,\rm GUT}}{\alpha_{\kappa}(M)}\right)^{\frac{1}{4}} \, 
\left(\frac{\alpha_{\overline{\lambda},\rm GUT}}{\alpha_{\overline{\lambda}}(M)}
\right)^{\frac{1}{4}}\,.\nonumber
\eea

We have presented the results from our numerical analysis in Table \ref{n4extended}. 
It is important to note that, already for moderate input values of the extra Yukawas
at the GUT scale, we reach the region of the experimentally measured $\alpha_3$.
Hence in models of this type it is no longer necessary to invoke excessively large Yukawa
couplings at the high-scale. For this reason we have restricted ourselves to a region
of parameter space where $\alpha_{\kappa,\rm GUT},\,\alpha_{\overline{\kappa},\rm GUT}, 
\,\alpha_{\lambda,\rm GUT},\,\alpha_{\overline{\lambda},\rm GUT} < 1$.
\begin{table}[ht]
\caption{Numerical results n=4 (two pairs of extra Yukawa couplings)}
\smallskip
\label{n4extended}
\centering
\begin{tabular}{|l|c|c|c|c|c|}
\hline
$M$(GeV) & $\alpha_{\rm GUT}$ & 
$\alpha_{\kappa,\rm GUT}=\alpha_{\overline{\kappa},\rm GUT} = 
\alpha_{\lambda,\rm GUT}=\alpha_{\overline{\lambda},\rm GUT}$ &
$y_t(M_{\rm GUT})$ & $2\pi/\alpha_3$   \\
\hline
\bf{1000}  & \bf{0.206} & \bf{0.228} &  \bf{2.25} & \bf{52.39}   \\
\hline
\textcolor{gray}{1000}& \textcolor{gray}{0.206}  & \bf{0.457} & 
\bf{3.38} & \bf{52.99}   \\
\hline
\textcolor{gray}{1000}& \textcolor{gray}{0.206}  & \bf{0.913} & 
\bf{5.20} & \bf{53.58}   \\
\hline
\end{tabular}
\end{table}

\section{Threshold corrections and higher-order effects} \label{thresh}

In this section we discuss issues related to low- and high-energy thresholds
and other higher-order corrections. The effect of the superpartner spectrum 
on the value of the strong coupling has been studied in detail in 
\cite{Langacker:1995fk,Carena:1993ag,Bagger:1995bw}. The analysis reveals that 
the low-energy thresholds potentially shift the predicted value of $\alpha_3$ 
by a significant amount. For instance, it is well-known that certain SUSY 
spectra with light gluinos can compensate for the detrimental two-loop effect 
discussed in Sect.~\ref{prediction} and therefore bring the $\alpha_3$-prediction 
in line with the experimental value (see e.g. \cite{Roszkowski:1995cn}). 
However, gluinos tend to be heavy in the simplest mediation scenarios and it 
generally requires a compensation of several mediation effects to make them light. 
A detailed study of concrete models realizing this possibility has recently 
appeared in \cite{Raby:2009sf}.

Another potentially important contribution arises from heavy particle 
thresholds. For example, from Eqs.~(\ref{dih}) or (\ref{dip}) it is clear 
that in order to shift the prediction for $2\pi/\alpha_3$ by several units 
the logarithms of the mass ratios $\,m^h_{X,Y}/m^h_3\,$ or $\,m^p_{X,Y}/m^p_3\,$ 
have to be several units themselves. In other words, the Higgs triplets have 
to be $\sim 10^2$ lighter than $M_{\rm GUT}$ and proton decay has to be avoided 
through some version of the missing partner mechanism (see \cite{Amsler:2008zzb}
for references). Of course, thresholds with larger numerical prefactors (and 
hence smaller required mass ratios) can arise in models with large GUT-scale 
representations (see, e.g., \cite{Langacker:1995fk,Ring:1995wc,Dutta:2007ai,
Nath:2006ut}). A similar enhancement can come from large multiplicities of heavy 
states (this has in particular been argued in the context of certain 
string-motivated models \cite{Ross:2004mi}).\footnote{For explicit orbifold 
constructions where the relevant corrections decouple from the string scale and 
their size could be checked straightforwardly see e.g. \cite{Blaszczyk:2009in}.} 
All of this clearly makes `GUT-scale thresholds' a viable explanation of 
the precise value of $\alpha_3$. 

The models considered in this paper offer an alternative solution to the 
two-loop $\alpha_3$-discrepancy. This solution differs conceptually from the 
aforementioned approaches as it does not rely on any type of threshold 
effects. It realizes a lower $\alpha_3$ value at the expense of relatively 
large extra Yukawa couplings. The latter do not require any additional 
SU(5)-breaking effect (beyond the doublet-triplet splitting, which is anyway
present in the MSSM). Nevertheless, precision at the high scale remains 
a critical issue and the rest of this section is devoted to its analysis. 

The extra multiplets supporting the extra Yukawas raise the value of the 
GUT-scale gauge coupling. It is then tempting to consider the extreme case 
of such scenarios: Grand unification at the strong-coupling point. Even 
more conservatively, one might want to drop the GUT-assumption altogether 
and to demand only that all three SM gauge factors become strongly coupled 
at the same energy scale \cite{Moroi:1993zj} (for early related work see
\cite{Parisi:1974cf}). The resulting high-scale error for the $\alpha_3$ 
prediction can be estimated using the familiar one-loop formula
\be 
\frac{4\pi}{\alpha_i(m_Z)}=\frac{4\pi}{\alpha_i(M_{\rm GUT})}+2b_i
\ln(M_{\rm GUT}/m_Z)\label{l1}
\,.
\ee
Naively, `strong coupling' means that the loop-expansion parameter 
$g^2/(16\pi^2)$ is of order one. This would imply that $4\pi/
\alpha_i(M_{\rm GUT})=1\pm {\cal O}(1)$ in Eq.~(\ref{l1}). The resulting error 
of the $\alpha_3$ prediction is rather small, around  $1\%$. However, 
because of the large number of flavors, the actual expansion parameter is 
in fact closer to $\alpha$ rather than $\alpha/(4\pi)$.\footnote{
There 
are at least two ways to see this: First, we focus on the contribution of 
all `flavors' to the one-loop $\beta$-function of the QCD coupling at high 
scales. The corresponding $\beta$-function coefficient is 
$2b_3^{\rm flavor}=2(n+6) \approx 20 \,\,$  ($n=4$ or $5$). This more than 
compensates for the suppression by $4\pi$, leaving us with a number close to 
$\alpha$ as the actual expansion parameter. Alternatively, we consider the 
one-loop contribution of the gluons/gluinos, $2b_3^{\rm color} = 18$. Once 
again, this is more than sufficient to cancel the factor of $4\pi$.
}
Being at strong coupling then means that $1/\alpha_i( M_{\rm GUT})=1\pm 
{\cal O}(1)$, which corresponds to a $\sim 10\%$ error of the $\alpha_3$ 
prediction. While this easily brings the 2-loop MSSM prediction for 
$\alpha_3$ in line with the data, it also makes any discussion of 2-loop 
effects obsolete: The error is simply too large.

For the purpose of the present paper, we adopt a different point of view: 
We assume that a model with true, calculable unification exists in principle 
and that the coupling strength is controlled by some high-scale holomorphic 
parameter (e.g. a string theory modulus). When talking about strongly-coupled 
unification, we assume that this model is realized in a region of its 
parameter space where $\alpha_{\rm GUT}\simeq {\cal O}(1)$.
 
To be more specific, consider the string-theoretic (heterotic) formula 
\cite{Nilles:1986cy}
\be 
f_i(S,T)=k_i S+\Delta_i(T)\label{stf}
\ee
for the gauge-kinetic functions of the three SM gauge groups. For the 
conventional embedding of $G_{SM}$ in SU(5) and SU(5) in E$_8$, the $k_i$
are unity, corresponding to tree-level unification. The crucial point 
is that the modulus governing the tree-level coupling strength (in this case
the dilaton superfield $S$) does not appear in the loop corrections. They 
depend on a set of different moduli which we collectively denote by $T$. The 
reason for this is basically the same as in field-theoretic arguments for 
1-loop running: holomorphicity and shift symmetry of Im($S$) \cite{
Novikov:1983uc,ArkaniHamed:1997mj}. Now, moving the modulus $S$ from its 
perturbative value to the region where the expansion parameter is 
${\cal O}(1)$, we see that the high-scale non-universal correction 
$\Delta_i(T)$ is not enhanced.\footnote{
The
dual situation, where the leading-order gauge coupling is governed by the
GUT-brane volume $T$ and corrections (related to a higher-dimension 
operator) depend on the dilaton $S$, arises in F-theory GUTs. These corrections 
tend to aggravate the two-loop discrepancy for $\alpha_3$ \cite{
Donagi:2008kj}, potentially making `our' Yukawa effect the more interesting
(see also \cite{Heckman:2010fh}).
}
The only potential danger comes from non-perturbative extra terms 
$\sim C_i\exp(-a_iS)$ which can appear on the r.h. side of Eq. (\ref{stf}). 
To keep such terms under control, we only have to assume that 
$\exp(-{\rm Re}S)\ll 1$ -- a much weaker requirement than $1/{\rm Re}S\ll 1$. 

A further important issue is the error which builds up along the RG 
trajectory from $m_Z$ to $M_{\rm GUT}$ as a result of using the two-loop 
instead of the full $\beta$-function for the gauge couplings. According to 
the previously discussed master formulae (cf. Eqs.~(\ref{hmf}) and 
(\ref{aip})) this error is identical to the error of the $\ln Z$ terms. At 
one loop (and focussing only on the gauge sector for simplicity) we have 
$\, d \ln Z / dt \sim \alpha\,$, which upon integration gives $\ln Z \sim 
\ln t$. It is clear that this leading order effect receives contributions 
from the entire integration range. We will now argue that higher-order 
corrections are UV dominated. To this end we recall that at two loops the 
RGE for a generic Z-factor has the form:
\be
\frac{d \ln Z}{dt} \, \sim \, \alpha + \alpha^2 \,, \label{schematic}
\ee
where $\alpha$ is the canonical (or physical) gauge coupling. From the 
anomaly relation Eq.~(\ref{anom}) we obtain schematically
\be
\alpha \sim \alpha_h + \alpha_h^2 \,\ln \alpha_h + \alpha_h^2 \,\ln Z \,
\sim \frac{1}{t} \, + \, \frac{\ln t}{t^2}\,,
\label{nonexact}
\ee
where we have only displayed the leading corrections. Note that we have 
used the (perturbatively exact) one-loop holomorphic gauge coupling 
$\alpha_h \sim 1/t$ as well as the fact that (at one loop) $\,\ln Z 
\sim \ln t \,$. Substituting the above relation in the two-loop 
RGE (\ref{schematic}) gives
\be
\ln Z \sim \int \frac{dt}{t} \, + \, \int \frac{\ln t}{t^2} \,dt\,,
\ee 
where terms $\sim 1/t^3\,$ or higher were neglected. The first term on the 
r.h. side gives the previously discussed leading order $\ln t$ effect. The 
second integral is UV dominated, i.e., this subleading effect can indeed be 
neglected at our level of accuracy. To be more precise, the corresponding 
correction is not enhanced by the parametrically large quantity $t\sim
\ln(M_{\rm GUT}/m_Z)$ or a log thereof. Hence, it corresponds to an 
${\cal O}(1)$ correction to $\ln Z$. This is equivalent to a multiplicative 
${\cal O}(1)$ uncertainty of the high-scale $Z$ factor, which we anyway have 
to accept in the absence of an explicit GUT model. The same argument goes 
through for contributions of even higher order.

To summarize, our strongly-coupled unification scenario is defined as follows: 
The unified gauge coupling is taken to be relatively large, 
$\alpha_{\rm GUT}\sim {\cal O}(1)$, while non-perturbative corrections are 
still under control. This may be the case because there is at least a (small) 
hierarchy of the type $\exp(-4\pi/\alpha_{\rm GUT})\ll 1$ or because the 
coefficients of such non-perturbative terms happen to be small. In such a 
setting, our 2-loop analysis of the strongly coupled model from Sect.~\ref{strong}
is meaningful and necessary.

We also emphasize that the aforementioned problems related to large and 
potentially incalculable GUT-scale corrections are automatically avoided in 
the models we considered in Sects.~\ref{perturbative} and \ref{extended}: 
All three gauge couplings remain within the perturbative domain throughout 
the entire energy range from $m_Z$ to $M_{\rm GUT}$.

\section{Conclusions}

In this paper we have shown that models with extra Yukawa couplings have a 
dramatic impact on the prediction for $\alpha_3$ at the weak scale. They 
can bring the two-loop prediction in line with experimental data without
appealing to large GUT-scale or weak-scale threshold corrections. This is 
an effect which has no analogue in the realm of MSSM physics -- even 
the top Yukawa coupling is negligible in this context. 

We introduced our main ideas using a simple model with strong GUT coupling. 
This model contains an extra $\bf{10}+\overline{\bf{10}}$ pair with top-like
Yukawa coupling to the MSSM Higgs doublets, together with further $\bf{5}+
\overline{\bf{5}}$ pairs making the GUT-scale gauge coupling strong. 
In this context, the main features and implications of our construction 
can be understood in a completely analytical manner. 

We then demonstrated that our promising initial results retain their 
validity in situations where the gauge couplings do not leave the 
perturbative domain. This more complete and partially numerical analysis 
revealed, among other things, that large input values for the extra Yukawa 
couplings at the GUT scale can lead to an almost perfect prediction for the 
electroweak-scale strong coupling.

We also tested the sensitivity of our results to extensions of the minimal
setting described above. In particular, the positive effect of the 
extra Yukawas on the unification prediction for $\alpha_3$ is significantly 
further enhanced for a complete vector-like extra generation with both up-type 
and down-type Yukawa couplings. In such models, perfect agreement with experiment 
is achieved without invoking excessively large values for the extra Yukawa or 
gauge couplings at the GUT scale. By contrast, introducing several copies of 
$\bf{10}+ \overline{\bf{10}}$ with corresponding Yukawa couplings leads to 
a weaker enhancement of the Yukawa effect. This can be traced to the 
increased scale $M$ at which the extra multiplets decouple. Such a raised 
decoupling scale, which is necessary to keep the GUT coupling moderate, 
partially compensates for the positive effect of further extra Yukawas.

The presence of additional vector-like matter (at comparatively low-energies) 
and of extra Yukawa couplings, which have large values in the ultraviolet, 
can clearly have a significant impact on the MSSM phenomenology. We expect 
that the enhanced Higgs $Z$ factors, which are at the heart of the effect 
we analyse, would in particular affect the values of the Higgs mass 
parameters $m_{H_U}^2$, $m_{H_D}^2$, $\mu^2$ and $B\mu$ which are expected in
any concrete SUSY breaking model. We note that effects on the Higgs mass bounds 
in related non-supersymmetric models have recently been analysed in 
\cite{Gogoladze:2010in}.

Finally, as we have discussed in some detail in our paper, the increased 
value of the GUT gauge coupling does not lead to a precision loss of 
the unification prediction for $\alpha_3$. The basic reason is that 
holomorphicity forbids the dominant higher-loop effects, which arise only 
via the non-holomorphic $Z$ factors of MSSM chiral multiplets. However,
non-perturbative GUT scale corrections clearly need to be controlled. This 
is easily possible in our setting, since it is not necessary to move to the 
actual strong-coupling point. 

{\it Note added:} As has been pointed out to A.H. after this paper was 
published, the positive effect of large extra Yukawa couplings on the 
$\alpha_s$ prediction had been noted early on in a somewhat different 
setting \cite{Babu:1996zv}.

\section*{Acknowledgments}

We would like to thank Yasunori Nomura for many helpful and detailed 
discussions and for his hospitality during the visit of A.H. to the Berkeley 
Center for Theoretical Physics. We are also grateful to M. Luty, T. Plehn
and M. Schmaltz for important discussions and comments.

\appendix

\section*{Appendix A: Gauge group Z-factors in the MSSM}

From the general formula (\ref{zf}) we obtain:
\bea
Z_{U(1)} &=& 
\left(\frac{\alpha_{\rm GUT}}{\alpha_1(m_Z)}\right)^{-\frac{2C_{F,1}}{b_1}} \, = 
\, 
\left(\frac{\alpha_{\rm GUT}}{\alpha_1(m_Z)}\right)^{-\frac{1}{22}} \, = 
\,0.96 \,\, \Rightarrow \,\, \ln Z_{U(1)} \,=\, -\, 0.04
\nonumber\\
Z_{SU(2)} &=& 
\left(\frac{\alpha_{\rm GUT}}{\alpha_2(m_Z)}\right)^{-\frac{2C_{F,2}}{b_2}} \, = \, 
\left(\frac{\alpha_{\rm GUT}}{\alpha_2(m_Z)}\right)^{-\frac{3}{2}} \,=\,0.75\,\, 
\Rightarrow \,\, \ln Z_{SU(2)} \,=\, -\, 0.29 \quad\quad\quad
\\
Z_{SU(3)} &=& 
\left(\frac{\alpha_{\rm GUT}}{\alpha_3(m_Z)}\right)^{-\frac{2C_{F,3}}{b_3}} \, = 
\, \left(\frac{\alpha_{\rm GUT}}{\alpha_3(m_Z)}\right)^{\frac{8}{9}} \, = 
\,0.39 \,\, \Rightarrow \,\, \ln Z_{SU(3)} \,=\, -\, 0.93\,\,.
\nonumber
\eea
Here we have used a one-loop GUT coupling $2\pi/\alpha_{\rm GUT}=2\pi\cdot 
24.3=153$. 
The quadratic Casimirs $C_{F,i}$ as well as the one-loop MSSM 
$\beta$-function coefficients are given by:
\be
2C_{F,i} = \left( \frac{3}{10}, \frac{3}{2}, \frac{8}{3} 
\right)\qquad\mbox{and}\qquad b_i = \left( \frac{33}{5}, \,1, \, -\,3 
\right)\,.
\ee

\section*{Appendix B: Effect of top in the MSSM}

In this appendix we analyze the effect of the top Yukawa coupling 
on the two-loop MSSM $\alpha_3$-prediction. We recall that in the
presence of Yukawa couplings the matter field $Z$-factors factorize
as $Z = Z^G \times Z^Y$, where, as before, the gauge part is given 
by Eqs.~(12) and (13). In order to obtain the Yukawa sector
contribution we start with the one-loop RGEs
\be
2\pi\, \, \frac{d \ln Z_{H_U}^Y}{dt} = - 3 \alpha_t\,,\qquad
2\pi\, \, \frac{d \ln Z_{U_3}^Y}{dt} = - 2 \alpha_t\,,\qquad
2\pi\, \, \frac{d \ln Z_{Q_3}^Y}{dt} = -   \alpha_t\,,\label{zrget}
\ee
where we have defined $\alpha_t = y_t^2/4\pi$, with $y_t$ being the 
\textit{canonical} top Yukawa coupling. The subscript `3' in `$U_3$'
and `$Q_3$' refers to the third generation quark-antiquark particles.
We also have the following one-loop equation for $\alpha_t$:
\be
2\pi \frac{d \ln \alpha_t}{dt} \, = \, 6\,\alpha_t \, - \, 
\frac{16}{3} \alpha_3 - 3 \alpha_2 - \frac{13}{15}\alpha_1\,.\label{trge}
\ee
Following \cite{AmelinoCamelia:1998tm}, we combine Eqs.~(\ref{zrget}) 
and (\ref{trge}), finding
\be
2\pi \frac{d \ln Z^Y_{H_U}}{dt} \, = \, - \, \frac{8}{3} 
\alpha_3 - \frac{3}{2}\alpha_2 - \frac{13}{30}\alpha_1 - 2\pi\frac{1}{2}
\frac{d \ln \alpha_t}{dt}\,.\label{difeq}
\ee
If we now express each of the three gauge couplings through their 
respective one-loop RGEs,
\be
\frac{d\ln \alpha_i^{-1}}{dt} = -\, \frac{b_i}{2\pi}\,\alpha_i\,,
\ee
we can integrate Eq.~(\ref{difeq}) analytically:
\be
Z^Y_{H_U}(m_Z) = \left(\,\frac{\alpha_{\rm GUT}}{\alpha_1(m_Z)} 
\right)^{\frac{13}{30\,b_1}} 
\left(\,\frac{\alpha_{\rm GUT}}{\alpha_2(m_Z)} 
\right)^{\frac{3}{2\,b_2}}
\left(\,\frac{\alpha_{\rm GUT}}{\alpha_3(m_Z)}\,
\right)^{\frac{8}{3\,b_3}}
\left(\,\frac{\alpha_t(M_{\rm GUT})}{\alpha_t(m_Z)}\right)^{\frac{1}
{2}}\,.\label{int}
\ee

As already mentioned, we focus on the moderately large $\tan 
\beta$ region, where $m_t=y_t v$ with $v=174$ GeV. Using the low-scale 
values $m_t=173$ GeV and $y_t(m_Z)=0.99$, we solve Eq.~(\ref{trge}) 
numerically, using the explicit one-loop formulae for $\alpha_i(t)$. 
The resulting high-scale value $y_t(M_{\rm GUT})=0.57$ is then used to 
obtain $\ln Z^Y_{H_U}(m_Z)=0.74$. Furthermore, Eqs.~(\ref{zrget}) 
imply $Z_{U_3}^Y=(Z^Y_{H_U})^{2/3}$ and $Z_{Q_3}^Y=(Z^Y_{H_U})^{1/3}$. 
Thus, we find
\be
\Delta_{\rm top}\left(\frac{2\pi}{\alpha_3}\right) =
\frac{3}{2}\ln Z_{Q_3}^Y-\frac{15}{14}\ln Z_{U_3}^Y 
+\frac{9}{14}\ln Z_{H_U}^Y = \frac{3}{7}\ln Z_{H_U}^Y = 0.32\,.
\ee
While, as is well known, this helps in lowering the $\alpha_3$ 
prediction, the effect is far too small.

\section*{Appendix C: Matter field Z-factors for n$\ge$1}

The numerical values of the wavefunction renormalization factors 
associated with the three gauge groups are listed in 
Table~\ref{zgauge}.

\begin{table}[ht]
\caption{Gauge group Z-factors}\label{zgauge}
\smallskip
\centering
\begin{tabular}{c c c c}
\hline\hline
\noalign{\smallskip}
$n$  &   $\ln Z_{SU(3)}$  &  $\ln Z_{SU(2)}$  &  $\ln 
Z_{U(1)}$ \\
\noalign{\smallskip}
\hline
1    &        -  1.06      &     -  0.33       &   - 0.04        \\
2    &        -  1.27      &     -  0.38       &   - 0.05        \\
3    &        -  1.63      &     -  0.46       &   - 0.06        \\
4    &        -  2.53      &     -  0.66       &   - 0.08        \\
4.45 &        -  3.87      &     -  0.93       &   - 0.11        \\
\end{tabular}               
\label{table:gauge}
\end{table}

Using these results it is then easy to calculate the two-loop corrections 
to $2\pi/\alpha_3$ arising from the different matter field $Z$-factors (cf. 
Table~\ref{zmatter}).

\begin{table}[ht]
\caption{Matter field contributions}\label{zmatter}
\smallskip
\centering
\begin{tabular}{c c c c c c c}
\noalign{\smallskip}
\hline\hline
\noalign{\smallskip}
$n$  &   $\frac{27}{14} \ln Z_L$  &  $ \frac{9}{2} \ln 
Z_Q $ & $ \,- \,\frac{9}{7} \ln Z_E $ & $ - 
\frac{45}{14} \,\ln Z_U $     &   $ - \frac{27}{14} \ln 
Z_D $ & $ \frac{9}{14} \ln Z_{H_U} +  \frac{9}{14} \ln Z_{H_D} $ \\
\noalign{\smallskip}
\hline
1    &   -0.72   &   -6.28   &   0.21   &   3.64  &  2.07   & -0.46  \\
2    &   -0.83   &   -7.46   &   0.26   &   4.37  &  2.49   & -0.54  \\
3    &   -1.01   &   -9.44   &   0.27   &   5.64  &  3.19   & -0.66  \\
4    &   -1.42   &  -14.40   &   0.41   &   8.59  &  4.95   & -0.94  \\
4.45 &   -2.00   &  -21.67   &   0.57   &  13.07  &  7.56   & -1.34  \\
\end{tabular}
\end{table}

\end{document}